# Phases of Superfluid Helium in Smooth Cylindrical pores

T.R. Prisk[1], N.C. Das[1], S.O. Diallo[2], G. Ehlers[2], A.A. Podlesnyak[2], N. Wada[3], S. Inagaki[4], and P.E. Sokol[1]

[1]*Indiana University, Department of Physics, Bloomington, IN, 47408, USA*
[2]*Spallation Neutron Source, Oak Ridge National Laboratory, Oak Ridge, TN, 37831, USA*
[3]*Nagoya University, Department of Physics, Chikusa-ku, Nagoya 464-8602, Japan*
[4]*Toyota Central R&D Laboratories, Inc. Yokomichi, Nagakute, Aichi 480-1192, Japan*

Two different microscopic phases have been observed via inelastic neutron scattering for superfluid helium confined within highly monodisperse, smooth, and unidimensional silica pores. When the helium forms a thin film on the pore walls, it supports a dramatically modified phonon-roton spectrum as well as a 2D surface roton. The energies of these modified phonon-roton modes are consistent with predictions for a dilute, low-density film, while the energy of the 2D surface roton corresponds to that of a dense film. The relatively smooth and weak surface potential of the substrate permits the formation of a film with large dilute regions and only small compressed regions. When the pores are saturated with liquid, the modified phonon-roton spectrum disappears, and bulk-like modes coexist with the 2D surface roton. Presumably, bulk-like liquid occupies the core volume of the pores and high density liquid layers are present at the liquid-solid interface. These findings clearly connect the nature of the excitations to the local density of the liquid.

Liquid helium infused within porous media has proved to be a playground for exploring the effects of confinement, disorder, and reduced dimensionality on quantum liquids[1-2]. Interesting and non-trivial changes occur to the physical properties of liquid $^4$He in confinement, including changes in critical exponents, thermodynamic functions, and transport properties. Neutron scattering studies have shown that the existence of a new microscopic excitation, classified as a 2D surface roton, is linked to the changes in the macroscopic properties of the confined superfluid[3]. However, the nature of this surface roton, and its relationship to the adsorption potential of the substrate, is not yet fully understood. In this paper, we report new inelastic neutron scattering measurements of the collective excitation spectrum of superfluid helium confined within nano-scale pores with walls known to be smoother than previously investigated materials. In very thin films, we observe a dramatically modified phonon-roton spectrum coexisting with the 2D surface roton. The energies of these modes are consistent with variational predictions for a *dilute* 2D film[4-6], while the energy of the 2D surface roton is consistent with a *dense* liquid. Upon filling of the pores, the dilute film disappears, the 2D surface roton strengths in intensity, and a bulk-like phonon-roton spectrum coexists with the 2D surface roton. Presumably, at full pore, bulk-like liquid occupies the core volume of the pores and compresses the surface layers near the liquid solid interface. These changes in the spectrum clearly connect the character of the excitations to the *local density* of the adsorbed superfluid. They also reveal the previously unknown relationship between atomic scale roughness of the substrate and the 2D surface rotons: surface roughness locally increases the effective surface adsorption potential, thereby compressing small regions of the adsorbed film and permitting only short-wavelength, roton-like excitations in these regions.

Inelastic neutron scattering has been extensively used as a direct probe of the microscopic dynamics of liquid helium confined in a variety of porous materials, including aerogel[7-9], Vycor[3,10], xerogel[11], Geltech[12], MCM-41[13], and MCM-48[14]. This body of work demonstrated that the excitation spectrum of superfluid $^4$He confined within nanometer scale pores shows nearly universal behavior: the confined liquid supports both bulk-like phonon-roton modes as well as two-dimensional layer modes existing within the high density fluid layers adjacent to the liquid-solid interface[15]. To date, these 2D modes, which have been classified as 2D surface rotons, have only been observed within the vicinity of the roton minimum $Q_R$. Using these 2D modes, with a 2D density of levels, and the 3D bulk like modes, Dimeo *et al*[3] successfully reproduced macroscopic, thermodynamic properties of the confined superfluid such as the superfluid density and heat capacity.

These previous neutron scattering studies have typically focused on materials with irregular pores and rough internal surfaces. Advances in the synthesis of mesoporous materials have produced materials with well-defined pores and smooth walls such as FSM-16, a folded sheet material with highly ordered pores that are nearly cylindrical and only a few nanometers in diameter[16]. Based on the mobility of adsorbed $^3$He studied by nuclear magnetic resonance spectroscopy, the surface adsorption potential is believed to be smoother than other porous glasses, including the MCM family of templated silica glasses[17-20]. Wada and his co-workers have performed torsional oscillator and heat capacity measurements of liquid helium adsorbed in FSM's as a function of temperature and pore filling[21-22], finding good evidence that adsorbed thin films of helium show 2D behavior. They argued that thin films of $^4$He in 28 Å FSM-16 pores undergo a finite-size Kosterlitz-Thouless phase transition in which vortex-antivortex pairing is influenced by strong confinement forces stemming from the restricted geometry[23]. The heat capacity[24] at milliKelvin temperatures $T$ goes like $T^2$, as expected from a phonon gas in reduced dimensions[25].

In this paper, we report inelastic neutron scattering measurements of the collective excitation spectrum of liquid $^4$He adsorbed within 28 Å FSM-16 pores. The powdered FSM-16 sample was synthesized using methods already described[16]. The sample was characterized using small-angle X-ray scattering and $N_2$ adsorption/desorption isotherms[26] at 77 K. The ordered pores of FSM-16 comprise a 2D triangular lattice structure and four clearly discernible Bragg reflections are observed, corresponding to a center-to-center pore spacing of 45 Å. The $N_2$ adsorption isotherm is Type IV[27], and has a steep capillary condensation branch with no detectable hysteresis. The pore size distribution, calculated from the capillary condensation branch using both the Pierce[28] and Barrett-Joyner-Halenda[29] methods, is sharply peaked near 28 Å and has a full-width at half-maximum of ~2 Å. The specific surface area determined by a Brunauer-Emmett-Teller (BET) plot[26] for pressures $0.05 < P/P_0 < 0.2$ is 1015 m$^2$/g.

The state of helium confined in FSM-16 has been extensively characterized by helium gas sorption isotherms at a series of different temperatures[30]. A $^4$He adsorption isotherm at 4.2 K on the FSM-16 sample used in this study confirms that adsorption proceeds by multilayer film growth and is in good agreement with previous results. A BET analysis for $0.05 < P/P_0 < 0.35$ gives monolayer coverage at $n_1 = 21.9$ mmol/g. It is likely that this first layer is an amorphous solid[31-33]. Saturated vapor pressure is achieved at roughly $n_{SVP} \sim 47$ mmol/g.

We performed inelastic neutron scattering measurements at the Spallation Neutron Source using the Cold Neutron Chopper Spectrometer[34-35] at Oak Ridge National Laboratory. The FSM-16 was contained in a cylindrical aluminum sample cell with mounted neodymium magnets to suppress its superconducting transition, and was attached to the cold plate of an Oxford Vericold dilution refrigerator. The thickness of the sample cell was chosen to allow for approximately 95% beam transmission in order to minimize the effects of multiple scattering. An incident neutron energy of 3.65 meV, which is below the Bragg edge of aluminum, was used to obtain a clean background signal with no strong spurious features. Gas loadings were dosed to the sample *in situ* at roughly 1 K.

Inelastic neutron scattering measures the dynamic structure factor $S(Q, E)$, which contains information about the density fluctuation spectrum of condensed matter systems[34,36]. The neutron scattering data distinguishes between three general classes of behavior observed in this system. For measurements taken at low fillings or within the normal liquid phase, we observe only substrate-like scattering and no well-defined collective excitations at any wavevector $Q$. As the filling is increased a distinct change in the scattering occurs. Figure 1a shows the scattering, $S(Q, E)$ at 49 mK for 37.7 mmol/g gas loading. As can be seen the scattering consists of a phonon-roton spectrum and has additional intensity below the roton minimum, implying the presence of a 2D surface roton. We will refer to these modes as dilute layer modes (DLM) and compressed layer modes (CLM), respectively. It is clear from inspection that the phonon velocity is reduced, the maxon is softened, and the curvature of roton region is altered relative to the bulk behavior when the pores are only partially filled with liquid. The shape of the scattering is insensitive to filling in the range from 33.4 to 37.7 mmol/g. When the filling is increased beyond 37.7 mmol/g a discontinuous change in the shape of the scattering is observed. Figure 1 (b) plots $S(Q, E)$ when the pores are fully saturated with superfluid helium. The phonon-roton spectrum occurs at nearly bulk energies along with the additional intensity at lower energy in the vicinity of the roton minimum. In both cases, no temperature dependence is observed below 1

K, and the well-defined modes altogether vanish as the temperature is raised far above the normal-superfluid transition line.

In order to extract the dispersion of the collective excitations, we performed least-squares fits of $S(Q, E)$ versus $E$ using the DAVE software package[37]. The scattering from the empty matrix was used as a background to be subtracted from the raw data. We found that, for $Q \leq 1.70$ Å$^{-1}$, the data could be fit by the sum of a linear background plus a single damped harmonic oscillator[34] (DHO) peak. For higher $Q$, where there is additional intensity at energies lower than the roton energy, it was necessary to include an additional Gaussian peak, representing the previously identified 2D surface roton or CLM[3].

The measured excitation energies at fillings of 33.4 and 37.7 mmol/g, where the helium forms a film less than 1 nm thick, are shown in Figure 2a. The observed roton spectrum exhibits considerable differences from the bulk spectrum[38], which is also shown in the figure. The phonon branch at low $Q$ has a lower slope than the bulk, indicating a reduced velocity of sound (~186 m/s), while the maxon is considerably lower in energy than the bulk value. The roton is shifted to lower momentum $Q_R = 1.782 \pm 0.006$ Å$^{-1}$, has a higher energy $\Delta = 8.97 \pm 0.01$ K and a higher effective mass $\mu = 1.80 \pm 0.08$ amu compared to the bulk values ($Q_R = 1.918 \pm 0.002$ Å$^{-1}$, $\Delta = 8.63 \pm 0.04$ K and $\mu = 0.64 \pm 0.03$ amu). The energies of the compressed layer modes cannot be determined with great precision at low fillings due to their low intensity. However, comparison of their excitation spectrum to the measured spectrum at higher fillings, shown as the solid line in Fig 2a, suggests there is little dependence on filling. As the filling is increased above 37.7 mmol/g, the excitation spectrum exhibits a qualitative change as shown in Figure 2b for fillings of 43.0 and 46.4 mmol/g. The phonon-roton-maxon spectrum becomes nearly identical to that of the bulk ($Q_R = 1.924 \pm 0.003$ Å$^{-1}$, $\Delta = 8.62 \pm 0.01$ K and $\mu = 0.78 \pm 0.04$ amu) and the compressed layer modes increase in intensity, but do not change in energy. There is no sign, within the experimental error of the measurement, of the modified phonon-roton excitation spectrum observed at lower fillings. There is also no indication of any dependence on filling, aside from the strength of the excitations. The compressed layer modes also exhibit a Landau type spectrum in the vicinity of the roton minimum with $Q_R = 1.94 \pm 0.01$ Å$^{-1}$, $\Delta = 6.93 \pm 0.08$ K and $\mu = 0.9 \pm 0.1$ amu, which are comparable to the values reported for other porous media[3,8-14].

The temperature dependence of the excitations below 1 K is weak, with little or no shifts in the peak positions at all fillings studied. When the confined liquid is in its normal state and far from its superfluid onset, there are no well-defined collective excitations. The bulk-like modes present at high filling have the same temperature dependence as true bulk liquid excitations, underscoring the independence of these excitations from the specific nature of the confining host. At $n$ = 43.0 mmol/g, the bulk-like roton minimum is softened to $\Delta_{BL}$ = 0.734 ± 0.001 meV when $T$ = 1500 mK.

The measurements are summarized on the phase diagram of $^4$He in FSM-16 in Figure 3, which also shows the normal-superfluid transition line disclosed by the torsional oscillator measurements of Ikegami et al[22]. There is a rapid crossover around ~40 mmol/g, illustrated on the phase diagram by a gray area. In general, our measurements of the temperature dependence spectrum agrees with the phase boundaries previously determined by torsional oscillator measurements, with a slight disagreement for one point close to that boundary. The inset to Figure 3 plots the film thickness $\delta$ and 2D isothermal compressibility $\kappa_T$ calculated from the $^4$He adsorption isotherm. The adsorbed film grows uniformly in statistical thickness until $n_f$ = 36.3 mmol/g, after which the film rapidly comes to saturate the pores. Clearly, the local maximum and minimum in $\kappa_T$ signal structural changes in the adsorbed film. The filling $n_f$ probably corresponds to the complete formation of a fluid second or third layer since the calculated film thickness at $n_f$ is ~7.5 Å. Quantum Monte Carlo (QMC) simulations[31] predict that helium atoms form concentric shells with approximately 3 Å spacing. Full pore can be estimated to be approximately 44 mmol/g by assuming that the confined liquid consists of the solid layer plus liquid at bulk density in the remaining core volume.

The evolution of the spectral weights of the bulk-like, dilute layer, and compressed layer rotons with pore filling are compared in Figure 4. The intensity of the compressed layer modes saturates as the pores are filled, which is due to the fact they inhabit the liquid layers adjacent to the interface and gain no further spectral weight after these layers are completed. The intensity of the DLM initially builds up as the thickness of the fluid film grows, but it drops to zero as the BL modes appear.

The evolution of the roton intensities as a function of filling suggests a simple physical interpretation. At low fillings, the adsorbed helium forms a thin fluid film on top of the solid layer directly adsorbed to the pore walls, and the fluid layers have dilute, low density regions and compressed, higher density regions. The excitation spectrum of the thin film therefore consists of what we have called dilute layer modes and compressed layer modes, both having a 2D density of levels. In previously studied porous media, only the compressed layer mode has been observed. Presumably the stronger surface potential in these other media is strong enough that the surface layer is compressed and leads to the compressed layer mode. Atomic scale roughness increases the effective strength of the substrate's adsorption potential[39]; the smooth surface potential in FSM-16 permits the formation of large dilute regions of the film and therefore the observation of dilute layer modes. At full pore, the liquid near the pore wall is compressed in the presence of bulk-like liquid at the pore center, and the excitation spectrum consists of bulk-like liquid occupying the core volume of the pores and high density layers adjacent to the liquid-solid interface. Bulk-like, 3D phonon-roton modes exist within the liquid occupying the core volume of the pores, while the 2D compressed layer modes inhabit the layers close to the surface. On this picture, the adsorbed superfluid undergoes a dimensional crossover from having only 2D excitations at low filling to a combination of 3D and 2D excitations at full pore.

Apaja and Krotscheck have argued that, at partial pore fillings, the energies of the bulk-like modes should reveal the microscopic properties of the bulk liquid under negative pressure[15]. The excitation spectrum of 2D planar helium at areal densities of 0.060 $\text{Å}^{-2}$ and 0.065 $\text{Å}^{-2}$ has been calculated by the variational method using both correlated basis functions[4] and shadow wavefunctions[5]. Based on QMC calculations[31], we expect that the aeral density of helium in the liquid layers close to the pore walls is approximately 0.070 $\text{Å}^{-2}$. The correlated basis functions and shadow wavefunctions semi-quantitatively predict the correct behavior: the phonon velocity $c$ is depressed to ~190 m/s, and the calculated $Q_R = 1.75$ $\text{Å}^{-1}$, both close to the observed values of 186 m/s and $1.782 \pm 0.006$ $\text{Å}^{-1}$. This is consistent with Wada's finding that the superfluid onset can be modeled as a 2D Kosterlitz-Thouless transition modified by the underlying silica substrate[23]. The spectrum of three-dimensional, bulk liquid has been calculated under negative pressure by path integral monte carlo methods[6,40] and shows the same general features. We have called the extended phonon-roton excitations observed in the adsorbed thin films dilute layer modes because their energies are consistent with those of a low density film.

In this paper, we report an inelastic neutron scattering studies of $^4$He confined within smooth, ordered pores only a few nanometers in diameter. When the adsorbed helium forms a fluid film only 1-2 atomic layers thick, the film chiefly consists of dilute, low density regions as well as some compressed, higher density regions. The excitation spectrum therefore consists of dilute layer modes and compressed layer modes. The properties of the dilute layer modes closely correspond to variational predictions for a low density 2D film. This finding buttresses the interpretation of torsional oscillator measurements as disclosing a modified Kosterlitz-Thouless phase transition at low fillings. At full pore, bulk-like liquid occupies the core volume of the pores and high-density layers are present near the interface with the solid. The excitation spectrum consists of 3D bulk-like modes and compressed layer rotons. Together, these findings clearly connect the character of the excitations with the local density of the liquid. It is unknown whether this dimensional change in the behavior of the system corresponds to a rapid crossover or a quantum phase transition, a topic of future inquiry.

This research at Oak Ridge National Laboratory's Spallation Neutron Source was sponsored by the Scientific User Facilities Division, Office of Basic Energy Science, U.S. Department of Energy. This report was prepared under award 70NANB5H1163 from the National Institute of Standards and Technology, U.S. Department of Commerce. The authors thank David Sprinkle, Lisa Deberr-Schmidt, and Saad Elorfi for their expert assistance and Matthew Bryan and Gerardo Ortiz for stimulating conversation.

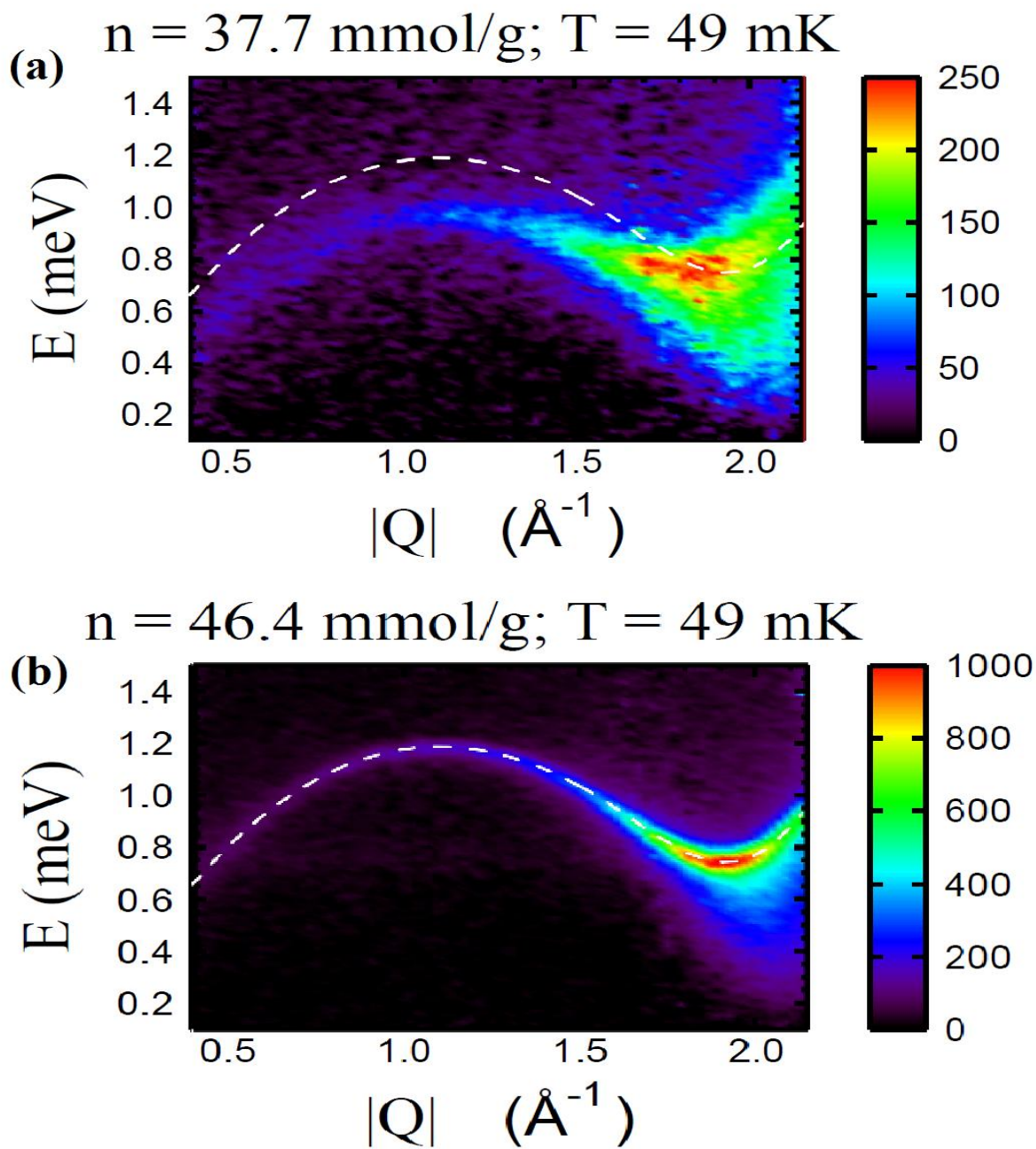

**Figure 1.** The dynamic structure factor $S(Q, E)$ at pore fillings 37.7 and 46.4 mmol/g at 49 mK, with the background subtracted. The dispersion of the bulk liquid is shown as a dashed white line.

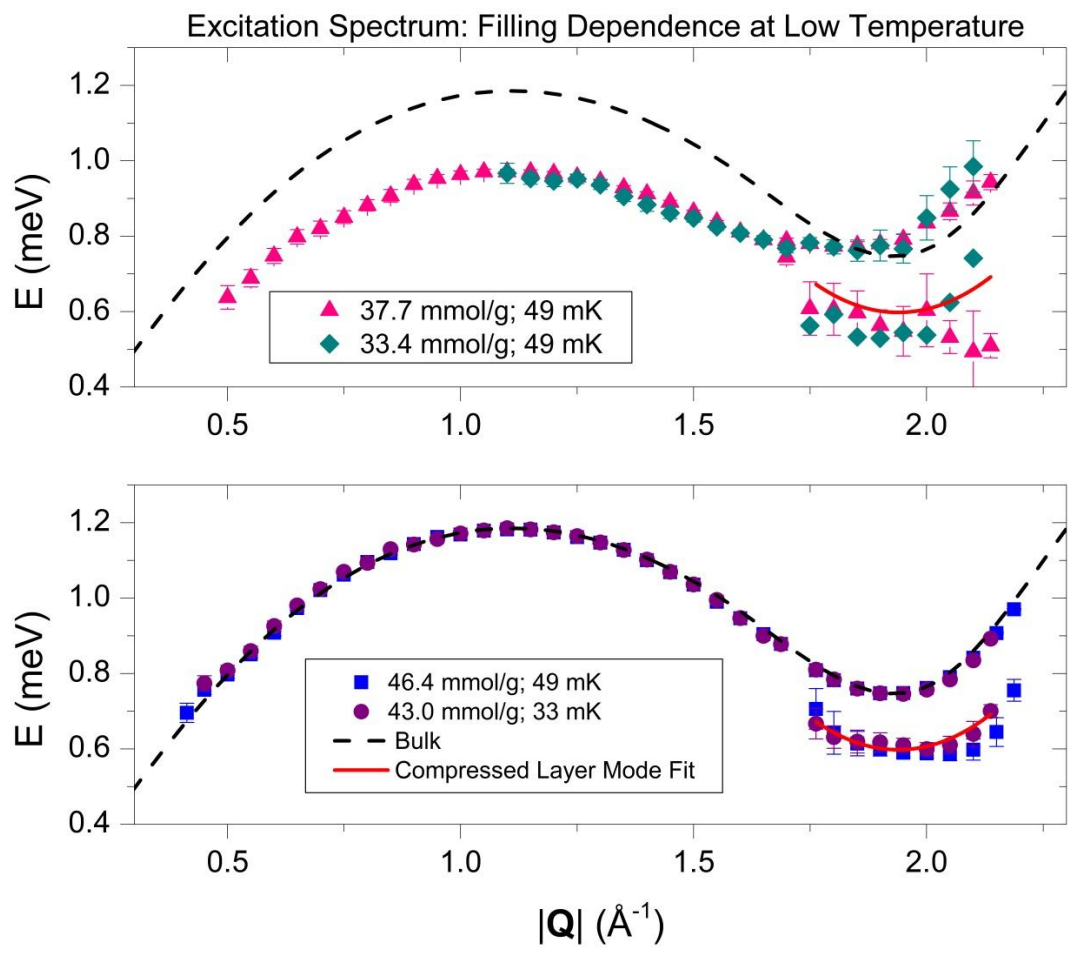

**Figure 2.** The filling dependence of the excitation spectrum is shown at temperatures near the base temperature of the cryostat. The dispersion of the bulk liquid is shown as a black dashed line.

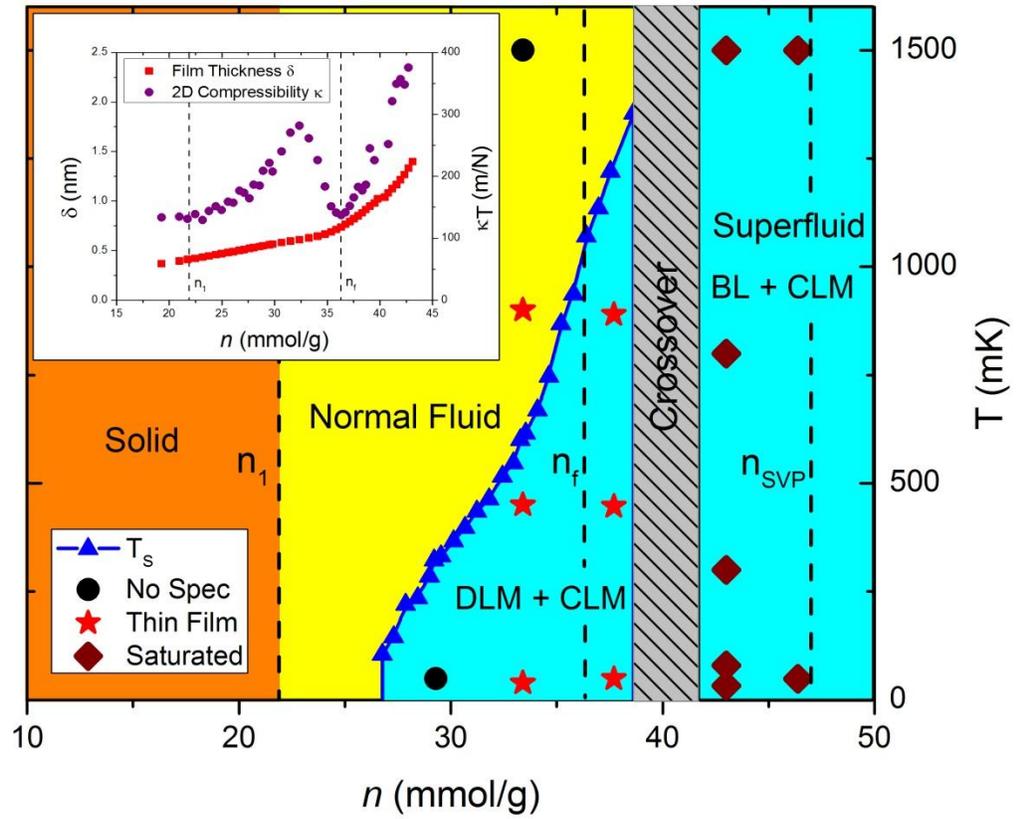

**Figure 3.** The temperature-filling phase diagram of $^4$He adsorbed in 28 Å FSM-16 pores.

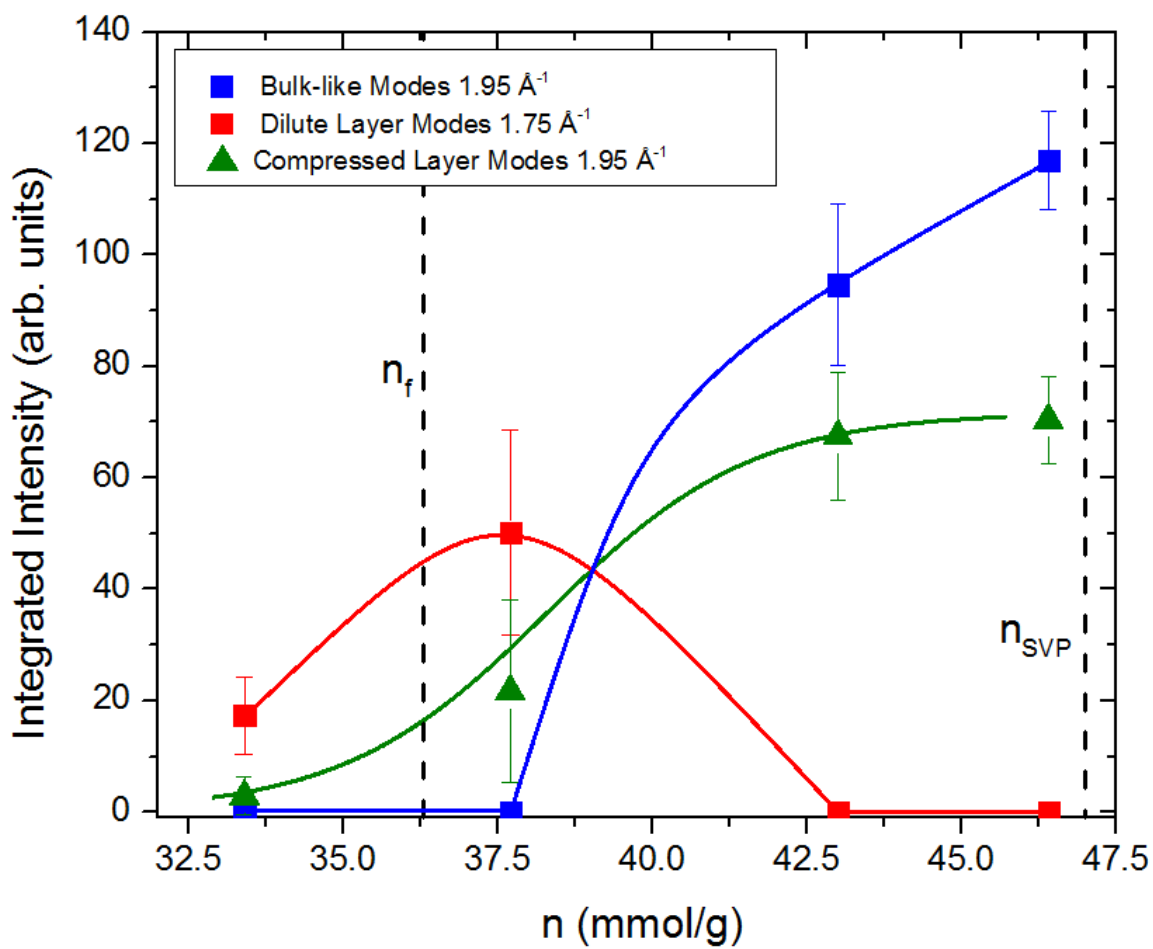

**Figure 4.** The evolution of the roton spectral weight with pore filling is shown here. The solid curves are guides to the eye.